\documentclass[prl,twocolumn]{revtex4}
\usepackage{graphicx,amssymb}

\newcommand{\beq}{\begin{equation}}
\newcommand{\eeq}{\end{equation}}
\newcommand{\beqa}{\begin{eqnarray}}
\newcommand{\eeqa}{\end{eqnarray}}
\newcommand{\cG}{{\cal G}}
\newcommand{\cH}{{\cal H}}
\newcommand{\vp}{\varphi}
\newcommand{\ve}{\varepsilon}
\newcommand{\sign}{{\rm sign}}
\newcommand{\bp}{{\bf p}}
\newcommand{\bg}{{\bf g}}
\newcommand{\tr}{{\bf\rm tr}}

\begin{document}

\title{Topological confinement in bilayer graphene}

\author{Ivar Martin$^1$, Ya.~M.~Blanter$^{2}$, and A. F. Morpurgo$^2$}
\affiliation{$^1$Theoretical Division, Los Alamos National
Laboratory, Los Alamos, New Mexico, 87544, USA\\
$^2$Kavli Institute of Nanoscience, Delft University of
Technology, Lorentzweg 1, 2628 CJ Delft, The Netherlands}

\date{\today}
\begin{abstract}
We study a new type of one-dimensional chiral states that can be created in
bilayer graphene (BLG) by electrostatic lateral confinement.  These states
appear on the domain walls separating insulating regions experiencing the
opposite gating polarity.  While the states are similar to conventional
solitonic zero-modes, their properties are defined by the unusual chiral BLG
quasiparticles, from which they derive.  The number of zero-mode branches is
fixed by the topological vacuum charge of the insulating BLG state.  We discuss
how these chiral states can manifest experimentally, and emphasize their
relevance for valleytronics.

\end{abstract}
\maketitle

Most condensed matters systems acquire gap in single electron excitation
spectrum at low temperatures.  Typically this happens as a result of
spontaneous symmetry breaking, as in the case of superconductors or charge- and
spin-density-wave materials.  The gap opens due to the interaction of electrons
with a slow bosonic mode representing the order parameter (OP). Yet, for
certain topologically non-trivial configurations of OP, zero-energy fermionic
states reemerge.  Examples are Andreev states that form at the domain walls in
superconductors~\cite{Andreev}, states in the superconducting vortex
cores~\cite{vortex}, and solitons in polyacethylene~\cite{schrieffer_su}.
Similarly, in cosmology, it has been suggested that our 3+1 dimensional space
with its extremely low elementary particle masses may represent a membrane or a
string in a higher-dimensional inhomogeneous Higgs vacuum \cite{RubakovS}.  The
{\em zero modes} may exhibit a number of interesting phenomena, including
fermion number fractionalization~\cite{Jackiw} and chiral
anomaly~\cite{Rubakov}.  In condensed matter, similarly to cosmology, the zero
modes originate from the Dirac equation, which emerges as an effective
(linearized) description of physics near the Fermi surface.

\begin{figure}[h]
\includegraphics[width=.9\columnwidth]{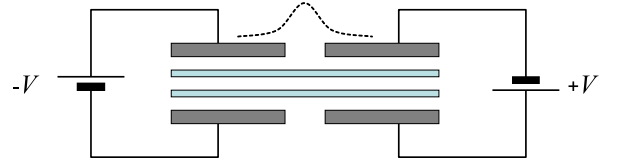}
\caption[] {Side view of gated bilayer graphene configuration with the voltage
kink.  The region where the interlayer voltage changes sign  (channel) supports
bands of chiral zero-modes (dashed line). The conventional (non-topological)
confinement would correspond to the same polarity of bias on both sides of the
channel.\label{fig:setup}}
\end{figure}

In this Letter we demonstrate that zero-modes can also emerge in
electrostatically-gated bilayer graphene (BLG) structures
(Fig.~\ref{fig:setup}).  These modes, however, are different from all the
examples from condensed matter and high-energy physics that we know of, as they
derive not from the Dirac fermions~\cite{Jackiw,Chamon}, but from the unusual
low-energy chiral modes of BLG~\cite{Falko}, which have quadratic dispersion
and zero gap between particle and hole bands.  When electrostatic bias $V$ is
applied between the layers, a gap of the magnitude $|V|$ opens between the
particle and hole bands \cite{McCann}.  The interlayer bias plays a role
analogous to an OP, but is externally {\em tunable}.  By applying {\em
inhomogeneous} bias one can spatially confine the low-energy states to the
regions with low gap, e.g. to a one-dimensional (1D) channel. However, in
addition to the conventional confinement by constant-polarity potential, in the
case of BLG, there is a possibility of {\em topological} confinement, with the
sign of the confining potential changing across the channel,
Fig.~\ref{fig:setup}.  We find that the topological confinement leads to
formation of 1D {\em chiral} zero-modes.  In each valley of graphene band
structure there are two such modes (per spin), both carrying electrons in the
same direction (opposite for the two valleys).  The robustness of the
zero-modes and their chiral nature are ensured by the topological structure of
the gapped bulk states.  These modes are likely to influence transport in BLG
with smooth potential disorder~\cite{voltageflucts}.  They may also have
implication for valleytronics~\cite{Beenakker}, as they can enable the
fabrication of valley filters and valves, which can be experimentally realized
with existing technology.

The low-energy (one-valley) bilayer Hamiltonian with bias $V(x)$
applied between the layers is~\cite{Falko,McCann}
\beqa\label{eq:Hfull}
H = \left(\begin{array}{cccc}
  -V(x)/2 & c\pi^\dag & 0 & 0 \\
  c\pi & -V(x)/2 & t_\perp & 0 \\
  0 & t_\perp & V(x)/2 & c\pi^\dag \\
  0 & 0 & c\pi & V(x)/2 \\
\end{array}\right),
\eeqa
where $c$ is the Fermi velocity, $\pi = p_x + ip_y$, $\pi^{\dag} = p_x - i
p_y$, and $t_\perp$ is the interlayer coupling.  The Hamiltonian acts in the
space of four-component wavefunctions, $(\psi_{A1}, \psi_{B1}, \psi_{A2},
\psi_{B2})$, where letters in subscript represent sublattice and numbers -- the
layer (we included here only one valley, assuming that $V(x)$ is smooth enough
not to cause intervalley transitions). This Hamiltonian provides a
good description of BLG as proven by recent experiments \cite{QHE}.
At low energies and constant gate
voltages, $V \ll t_{\perp}$, the Hamiltonian can be further
reduced~\cite{guinea},
\beqa\label{eq:tH}
\tilde H = \left( \begin{array}{cc}
  -\frac{V}{2}\left(1 - \frac{c^2 p^2}{t_\perp^2}\right) & -\frac{{c^2\pi^\dag}^2}{t_\perp} \\
  -\frac{{c^2\pi}^2}{t_\perp} & \frac{V}{2}\left(1 - \frac{c^2 p^2}{t_\perp^2}\right) \\
\end{array}\right).
\eeqa
Here $p = \sqrt{p_x^2 + p_y^2}$.  The remaining components of the wavefunction
have predominantly $A1$ and $B2$ character.  The spectrum of the Hamiltonian is
\beq
E^2 =\frac{V^2}{4}\left(1 - \frac{c^2 p^2}{t_\perp^2}\right)^2 + \frac{c^4
p^4}{t_\perp^2}.
\eeq
At finite bias, the spectrum has a gap which reaches minimum $|V|$ on the
circle $c p \approx V/2$.  However, as long as $V < t_\perp$, the $p^2$
correction in the first term can be neglected, simplifying spectrum to $E^2 =
V^2/4 + c^4p^4/t_\perp^2$.  The corresponding term can be also neglected in the
diagonal part of $\tilde H$.  As a result, in the Hamiltonian (\ref{eq:tH}),
the $V$ terms and the momentum terms are decoupled.  Thus, the transition to
the position-dependent potential $V({x})$ corresponds to reinstating the
momenta $p_x$ and $p_y$ as differential operators, finally leading to a
dimensionless quasiclassical Hamiltonian \cite{fnote},
\beq\label{eq:Hqc}
\cH_{qc} = -\vp(x)\sigma_z - (p_x^2 - p_y^2)\sigma_x - 2p_x p_y\sigma_y \equiv
 \bg(\bp,x)\cdot {\bf \sigma},
\eeq
where we defined $\varphi(x) = V(x)t_\perp a^2/(2 c^2)$, and momenta are
measured in units of inverse lattice constant $a$.

For constant $\vp(x)$ the spectrum has the form $E =
\pm\sqrt{\Delta^2 + \epsilon(p)^2}$, which is similar to the spectra of
condensed matter systems with off-diagonal long range order; the gate voltage
$\vp$ plays the role of the order parameter $\Delta$.  It is well known that in
such systems with non-trivial topological structure of $\Delta({\bf r})$,
low-energy fermionic modes can emerge.  Motivated by this analogy, we study
related inhomogeneous configurations of $\vp(x)$.

From Eq.~(\ref{eq:Hqc}), the corresponding wave equation is
\beqa
-\varphi(x) u + (\partial_x + p_y)^2 v = \ve u,\label{eq:SE1}\\
\varphi(x) v + (\partial_x - p_y)^2 u = \ve v.\label{eq:SE2}
\eeqa
It possesses a number of symmetries for a general antisymmetric
potential profile, $\vp(-x) = -\vp(x)$. It is easy to see that in this case for
$v(x) = \pm u(-x)$ the two equations (\ref{eq:SE1}) and (\ref{eq:SE2}) reduce
to one.  Therefore one can solve two problems separately, for $\Psi = [u(x),
u(-x)]$ and $\Phi = [w(x), -w(-x)]$. Furthermore, after we find the eigenvalues
and eigenvectors of the first problem for some value of $p_y$, $\Psi_{p_y}^n =
[u_{p_y}^n(x), u_{p_y}^n(-x)]$ and $\ve_{p_y}^n$, the other solution can be
obtained as $\Phi_{p_y}^n = [-u_{-p_y}^n(-x), u_{-p_y}^n(x)]$ with eigenvalues
$(-\ve_{-p_y}^n)$.  Thus in the $p_y-\ve$ plane the dispersion of $\Phi$ is
obtained from dispersion of $\Psi$ by inversion relative to the point $(0,0)$.
A related, useful symmetry has to do with the behavior of the spectrum  of Eqs.
(\ref{eq:SE1}) and (\ref{eq:SE2}) under $p_y\rightarrow -p_y$. Easy to see that
under this transformation $(u, v) \rightarrow (v, -u)$ and $\ve \rightarrow
-\ve$.

Unlike the 1D Dirac equation, which typically describes the domain-wall
zero-modes \cite{Jackiw}, Eqs. (\ref{eq:SE1}) and (\ref{eq:SE2}) are not easily
solvable analytically for a general profile $\varphi(x)$.  We therefore first
analytically study the simple case of step-like potential $\vp(x) = \vp_0\,
\sign(x)$, and then perform numerical solution of Eqs.~(\ref{eq:SE1}) and
(\ref{eq:SE2}) for a smooth $\varphi(x)$.  As we will see, only the details of
the zero-mode dispersion $\ve(p_y)$ depend on the exact profile of $\vp(x)$.
This is a manifestation of the topological nature of these states.

{\it Step kink.}  We first consider a step-like kink, $\vp(x) = \vp_0
\sign(x)$.  In this case, both for positive and negative $x$ the potential is
constant, and the solution of the wave equation in these regions is
$\Psi\propto e^{-\lambda x}$.  For the intragap states, i.e. those with $|\ve|
< \vp_0$,
\beqa\label{eq:lam}
\lambda &=& \pm\alpha \pm i\beta,
\eeqa
where $\alpha(\beta) = 2^{-1/2}[(p_y^4 +\vp_0^2 - \ve^2)^{1/2} +(-)
p_y^2]^{1/2}$.  For $x<0$ one should only keep $\lambda^<_{1,2} = \alpha \pm i
\beta$ and for $x>0$ only $\lambda^{<}_{1,2} = -\alpha \pm i \beta$, with the
corresponding wavefunctions of the form
\beqa
u^{\gtrless}(x) = u^{\gtrless}_1 \exp({-\lambda^{\gtrless}_1 x}) +
u^{\gtrless}_2 \exp({-\lambda^{\gtrless}_2 x}).\label{u1}
\eeqa
At $x = 0$ the left and right solutions have to be matched. From the structure
of the wave equation it is clear that the wavefunction and its first derivative
are continuous across $x = 0$, while the second and higher derivatives are not.
Considering for concreteness the states of the form $\Psi = [u(x), u(-x)]$, the
matching conditions are
\begin{eqnarray}
u^>&= & u^<\\
\partial_x u^>& =& \partial_x u^<\\
\partial_x^2 u^>& =&\partial_x^2 u^< -2\vp_0 u\\
\partial_x^3 u^>& =&\partial_x^3 u^< + 2p_y(\partial_x^2 u^> -\partial_x^2
u^<) + 2\vp_0 \partial_x u.
\eeqa
The third equation is obtained by subtracting Eq.~(\ref{eq:SE1}) at $x= -0$
from  itself at $x = +0$ and using $v(0) = u(0)$.  The forth equation is
obtained in the same manner but after first differentiating Eq.~(\ref{eq:SE1})
over $x$ and using that $\partial_x v(0) = -\partial_x u(0)$.  Substituting now
the general solutions Eqs.~(\ref{u1}), we obtain a homogeneous system of 4
equations with 4 unknowns which only has a non-trivial solution if its
determinant equals zero.  Given the form of $\lambda$, Eq.~(\ref{eq:lam}), this
condition is equivalent to
\beq
4\alpha^2(\alpha^2+\beta^2) + 4p_y\vp_0\alpha - \vp_0^2 = 0.
\eeq
Near zero energy, this equation has a solution only for $p_y <0$.  For $\vp_0
>0$, we obtain the dispersion
\beq\label{eq:disp}
p_y = -({\ve +\vp_0/\sqrt{2}})/{(\ve + \vp_0\sqrt{2})^{1/2}}
\eeq
(we analyzed the wave equation near $\ve = 0$ to remove a spurious
branch with $\ve \rightarrow -\ve$). The zero-energy solution obtains at $p_0 =
-\sqrt{\vp_0}/2^{3/4}$. The other branch corresponding to wavefunction $\Phi =
[-u_{-p_y}(-x), u_{-p_y}(x)]$ has, as discussed above, the inverted dispersion,
\beq \label{eq:disp-otherK}
p_y = ({-\ve +\vp_0/\sqrt{2}})/{(-\ve + \vp_0\sqrt{2})^{1/2}}.
\eeq
Notice that both solutions have negative velocity near $\ve = 0$. This seems to
imply time-reversal symmetry breaking.  However, for the second graphene
valley, the velocity is positive and thus the symmetry is reinstated.
Therefore, the zero-modes are chiral: i.e. if we define pseudospin-1/2, with
$S_y$ corresponding to the valley index, {\em all} zero-modes have definite
sign of $p_y S_y$.  It is also easy to see that on the anti-kink ($\vp_0 <0$)
the dispersion is flipped, $\ve(p_y) \rightarrow -\ve(p_y)$.

Note that the decay length of the wave function of the topologically confined
states is of the order of ${a/\sqrt{\vp_0}\approx a t/\sqrt{V t_\perp}} \gg a$
at low energies.  Therefore, our solution is consistent with the
long-wavelength expansion (\ref{eq:Hqc}) used for the description of the
system. For kinks wider than $a/\sqrt{\vp_0}$ one expects {\em quantitative}
deviations, which we indeed find in the direct numerical solution.

\begin{figure}[h]
\includegraphics[width=.8\columnwidth]{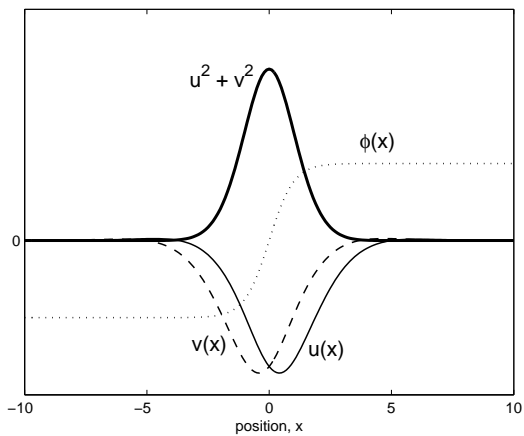}
\caption[] {Spatial structure of the confined state obtained numerically from
Eqs.~(\ref{eq:SE1}) and (\ref{eq:SE2}) for particular value $p_y = -1/2^{3/4}$.
The energy of the state is $\ve = 0.1140$. Shown are the components $u$ and $v$
of the wavefunction (thin and dashed lines, respectively), the probability
density $|\Psi|^2 = u^2 + v^2$ (thick line), and the potential profile, taken
here as $\vp(x) = \tanh(x)$.  Note the symmetry, $u(x) = v(-x)$.
 \label{fig:soli}}
\end{figure}

{\it Numerical solution.} The wave equations (\ref{eq:SE1}) and
(\ref{eq:SE2}) can be solved for an arbitrary potential profile $\vp(x)$
numerically using the finite differences method.  In Figure \ref{fig:soli} we
show an example of an electronic state localized at the kink of the potential
$\vp = \tanh(x/\ell)$ for $\ell =1$ and $p_y = -1/2^{3/4}$.  For step-like
potential this value of $p_y$ would correspond to a zero-energy state.  For the
rather smooth potential used here, the lowest energy is finite, $\ve = 0.1140$,
but indeed much smaller than the reference gap, $\vp_0 = 1$. The state has the
expected symmetry, $v(x) = u(-x)$.  The wavefunction has a slightly oscillatory
and  decaying behavior, consistent with the complex (non-real) values of
$\lambda$'s.  A symmetry-related low energy state $\Phi = (-v, u)$ occurs at
$p_y = +1/2^{3/4}$.

\begin{figure}[h]
\includegraphics[width=.49\columnwidth, height = 3cm]{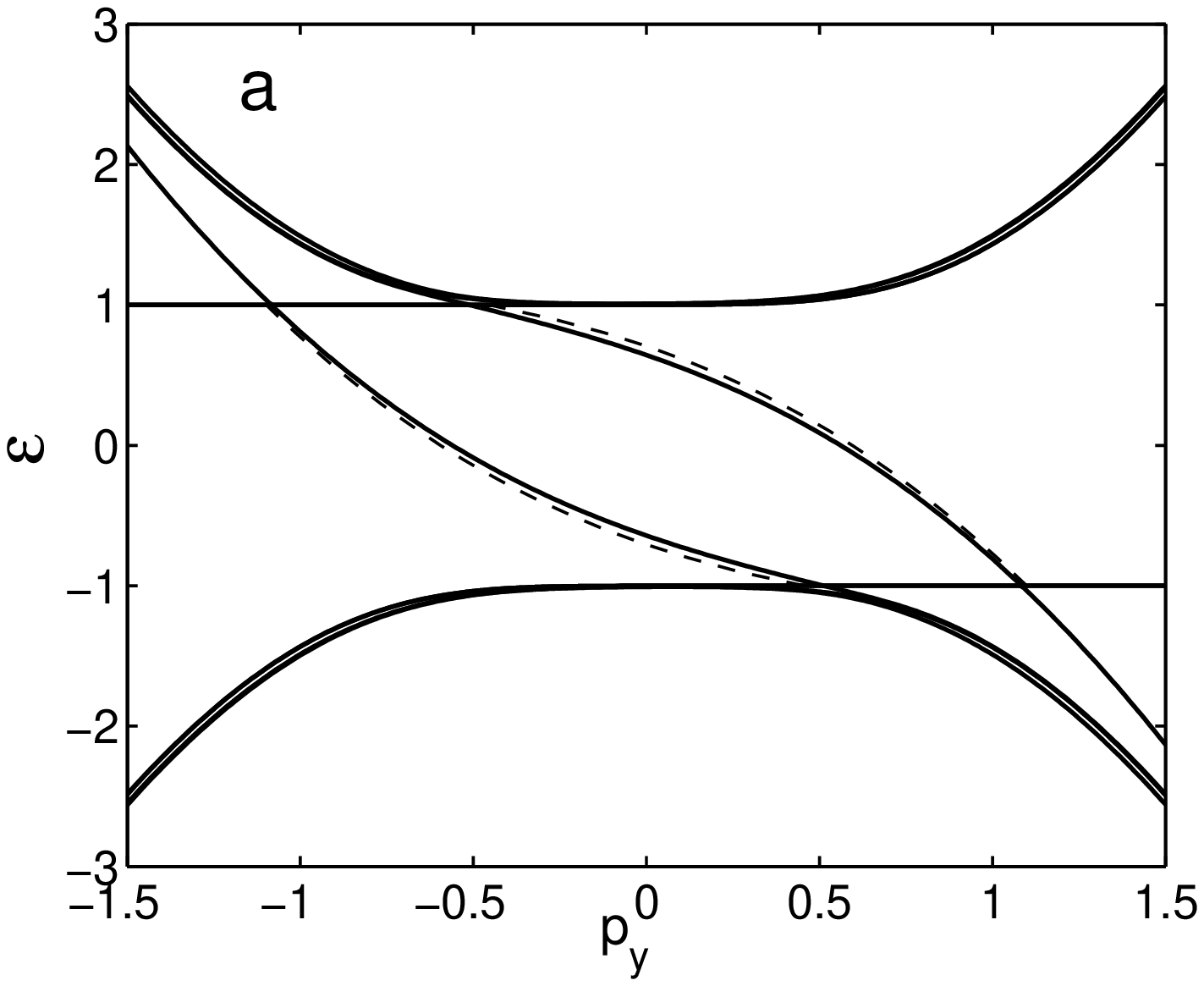}
\includegraphics[width=.49\columnwidth, height = 3cm]{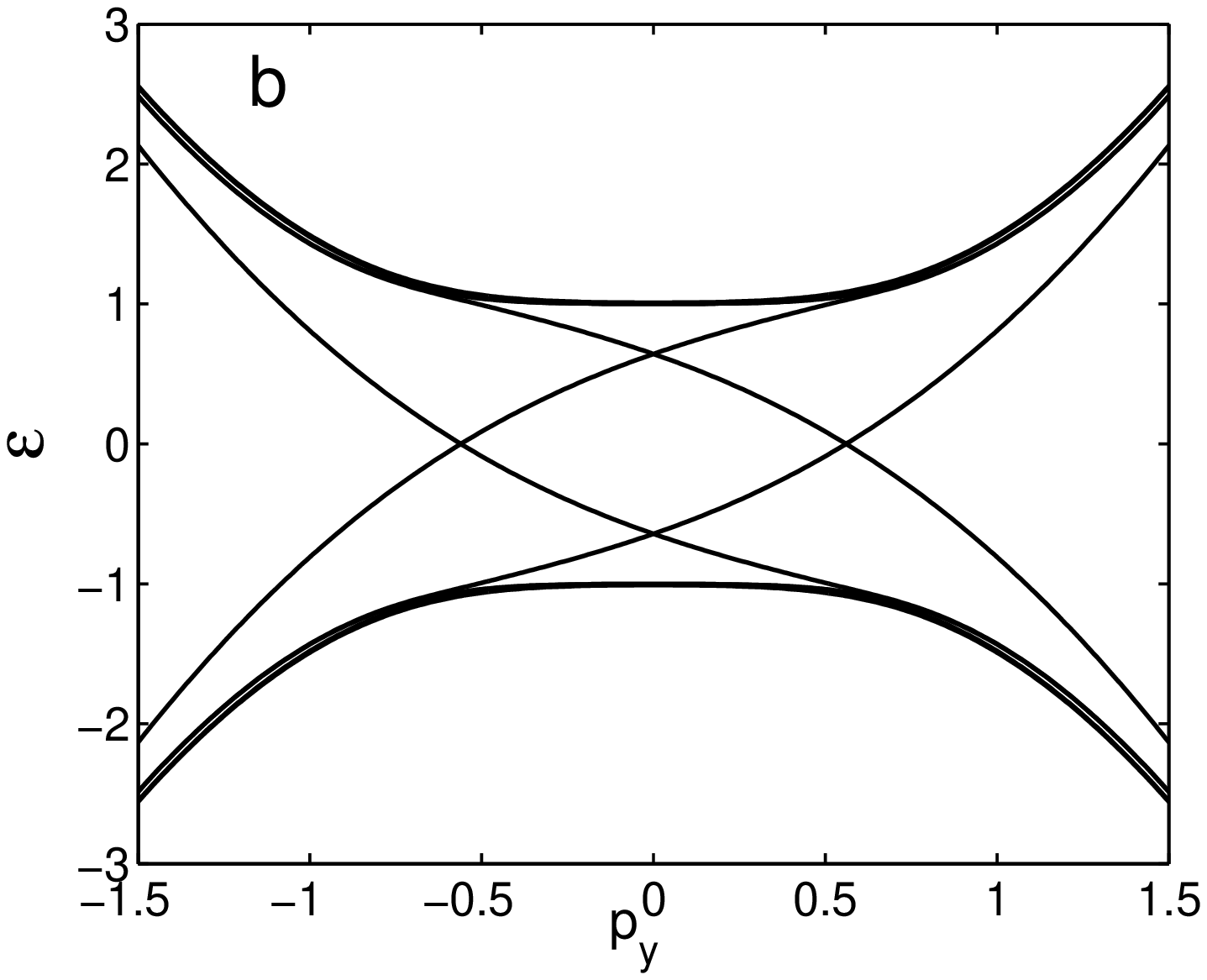}
\vspace{3 mm}
\includegraphics[width=.49\columnwidth, height = 3cm]{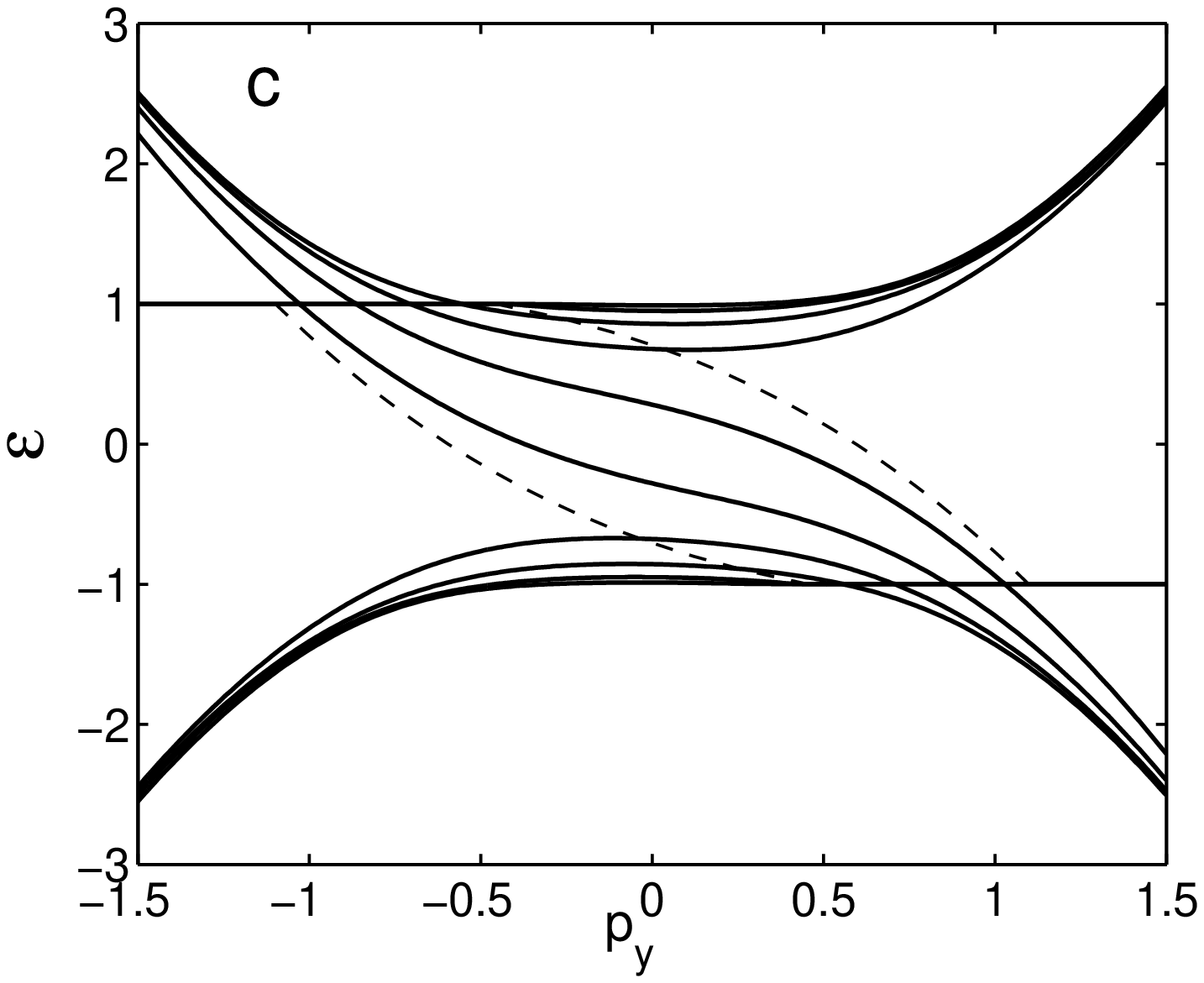}
\includegraphics[width=.49\columnwidth, height = 3cm]{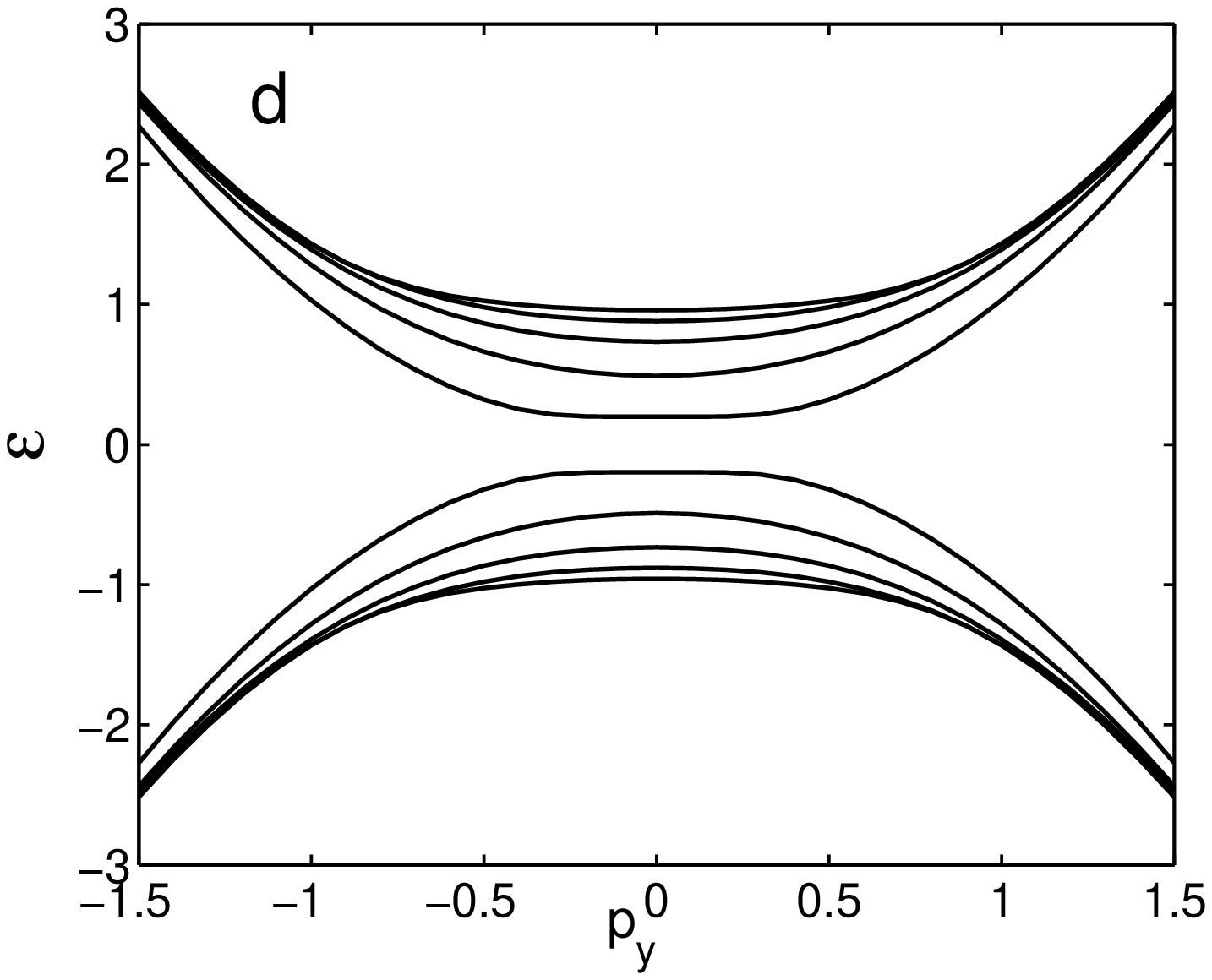}
\caption[] {Energy level structure in the presence of non-uniform interlayer
bias $\vp(x)$.  The low-energy states in (a) and (c) are localized on the
single kink of profile $\vp(x)= \tanh(x/\ell)$ with $\ell = 0.5$ and $\ell =
4$, respectively.  For (a) and (c) we used open boundary conditions, $\Psi(\pm
20) = 0$, which caused flat bands at $\ve = \pm 1$.  The dashed lines are the
analytical expressions for the intragap state dispersion, Eq.~(\ref{eq:disp}).
(b): Same as (a) but with kink and anti-kink at $x = \pm 10$ and periodic
boundary conditions.  (d): Non-topological confinement for potential profile
$\vp(x)= \tanh^2(x/4)$.  Note the absence of chiral zero-modes.
\label{fig:bands}}
\end{figure}

In Figures \ref{fig:bands}a--c we show the intra-gap state dispersions
$\ve(p_y)$ for kinks of various widths.  Note that for the narrow kink (panels
a and b) there are only two intragap states, while for the wide kink (panel c)
more low-energy states appear below the gap edge. In all cases, there are only
two (chiral) zero-modes per kink per valley per spin regardless of the width.
As expected, the analytical result, Eqs.~(\ref{eq:disp}) and
(\ref{eq:disp-otherK}), fits very well for the narrow kink.  For comparison,
panel d shows the dispersion relation for states in a 1D channel defined by
``conventional'' confinement. In this case no chiral zero modes are present,
even though there are states below the gap edge.

{\it Charge of the kink.}  It is well know that solitons in 1D systems can
carry charge, which can be either rational or irrational
\cite{wilczek,rice_mele,jackiw_semenov,kivelson,Chamon}. Similarly, in our
case, one might expect that the kink can carry charge.  Let us demonstrate that
contrary to this expectation the domain wall charge density is zero. Using the
symmetries of the problem, the density in the presence of a kink is
proportional to
\beqa
\sum_n\int_{-\infty}^{\infty}{|\Psi^n_{p_y}(x)|^2} n_F(\ve^n_{p_y})=
\sum_n\int_{0}^{\infty}{dp_y |\Psi^n_{p_y}(x)|^2}.\nonumber
\eeqa
By completeness property of the wavefunction, this expression is an
$x$-independent constant proportional to the momentum (energy) cut-off.

{\it Topological considerations.}  Above we found that in the presence of a
kink there are two zero-modes per graphene valley per spin. We now demonstrate
how the appearance of these modes can be understood from topological
considerations \cite{volovik}. The number of zero-modes, or more precisely the
number of zero-energy left-movers minus the number of zero-energy right movers
is related to the topological charge $N_3$ of the Fermi point located in the
extended quasiclassical 3D momentum-real space $(\bp, x)$.  In terms of the
quasi-classical Green function, $\cG_{qc}= (ip_0 - \cH_{qc})^{-1}$,
\beqa
N_3 = \frac{1}{24\pi^2} e_{ijkl} \tr \int_{\sigma_3}{dS^l \cG_{qc}\partial_i
\cG_{qc}^{-1} \cG_{qc}\partial_j \cG_{qc}^{-1} \cG_{qc}\partial_k
\cG_{qc}^{-1}},\nonumber
\eeqa
where integration is to be performed over an arbitrary 3D surface $\sigma_3$
enclosing the Fermi point  $(p_0, \bp, x) = (0,{\bf 0}, 0)$.  The charge $N_3$
is the degree of the mapping of $\sigma_3$ onto a manifold corresponding to the
$\cG_{qc}$.  Given the form of $\cH_{qc}$, this manifold is equivalent to a 3D
sphere, and thus the mapping belongs to non-trivial $\pi_3(S^3)$ homotopy
group.  In evaluating $N_3$, two choices of $\sigma_3$ are particularly useful:
(a) infinite planes $p_y = \pm p_y^0$ and (b) infinite planes at $x = \pm x^0$
(assuming that the planes are sufficiently close, the flux contributing to
$N_3$ but not crossing these planes is negligible).  The representation (a) is
equivalent to the quasiclassical expansion of the difference of spectral
asymmetry functions, $\nu = \nu(p_y^0) - \nu(-p_y^0)$, where
\beqa
\nu(p_y) = \tr\int\frac{dp_0}{2\pi i} \cG \partial_{p_0} \cG^{-1} =
-\frac{1}{2}\sum_n\sign \,\ve^n_{p_y}.\nonumber
\eeqa
$\nu$ counts the number of dispersion branches that cross zero energy from
above on the interval $[-p_y^0,p_y^0]$.  In the representation (b), $N_3$ is
nothing but the difference of the vacuum topological charges of the insulating
states to the right and to the left of the domain wall, $\tilde N_3(x^0) -
\tilde N_3(-x^0)$.  Given the form of $\cH_{qc}$, Eq.~(\ref{eq:Hqc}),
\beqa
\tilde N_3(x_0) = \frac{1}{4\pi}\int  dp_x dp_y \frac{1}{|\bg|^3}
\bg\cdot[\partial_{p_x}\bg\times\partial_{p_y} \bg],\nonumber
\eeqa
from which one easily finds $\tilde N_3(x_0) = \sign[\vp(x^0)]$.  Thus, for
$\vp(+\infty) >0$, we obtain $N_3 = \tilde N_3(x_0) - \tilde N_3(-x_0) = \nu =
2$, consistent with our earlier finding that there are two branches per kink
per valley per spin that cross zero energy with negative velocity.

Having established the presence of one-dimensional chiral zero-modes
originating from topological confinement, we briefly discuss their relevance to
one area of current interest, namely ``valleytronics,'' which attempts to
utilize the valley degree of freedom to achieve new electronic
functionality~\cite{Beenakker}.  As we have shown above (e.g. see Fig.
\ref{fig:bands}a,c), a topologically confined 1D channel contains zero modes,
with the direction of propagation determined by the valley.  Thus, a {\em
valley filter} is realized when a voltage difference is applied along the kink
-- only electrons in one of the valleys will carry the current and the other
valley is ``filtered out."  It is also straightforward to realize a {\em valley
valve} by simply ``connecting" two filters in series.  When the polarity of two
filters is the same, the current can pass through; for opposite polarities the
current is blocked.  We note that earlier proposals~\cite{Beenakker,Recher} for
valley filters and valves in graphene monolayers relied on perfect zig-zag
edges, making their practical realization very challenging.  On the other hand,
the technology needed for valley filter and valve using the
topologically-confined channels is much less demanding: A smooth 1D channel can
easily be realized away from the edges of the graphene sample, thus decreasing
substantially the possibility of intervalley scattering.  Some of the
technology--the opening of a gap in a bilayer using gate electrodes--has
already been demonstrated experimentally~\cite{topgates}. A more detailed
discussion of valley filtering induced by topological confinement will be
presented elsewhere \cite{prep}.

We thank M. Chertkov, V. Chernyak, and J. Zaanen for discussions and
suggestions.  This work was supported by the U.S. DOE (IM), and Dutch NWO
(A.F.M.).


\begin{thebibliography}{99}
\bibitem{fnote} We have verified by a direct numerical
calculation that such reduced Hamiltonian well approximates the full
Hamiltonian (\ref{eq:Hfull}).
\end{thebibliography}
\end{document}